\documentclass[twocolumn,showpacs]{revtex4}

\usepackage{graphicx}
\usepackage{dcolumn}
\usepackage{amsmath}

\begin{document}

\title{Mott transition in cuprate high-temperature superconductors
}


\author{Takashi Yanagisawa$^a$ and Mitake Miyazaki$^b$}

\affiliation{$^a$Electronics and Photonics Research Institute,
National Institute of Advanced Industrial Science and Technology (AIST),
Tsukuba Central 2, 1-1-1 Umezono, Tsukuba 305-8568, Japan\\
$^b$Hakodate National College of Technology, 14-1 Tokura, Hakodate,
Hokkaido 042-8501, Japan
}

\begin{abstract}
In this study, we investigate the metal-insulator transition of charge
transfer type in high-temperature cuprates.
We first show that we must introduce a new band parameter in the three-band
d-p model to reproduce the Fermi surface of high temperature cuprates such as
BSCCO, YBCO and Hg1201.
We present a new wave function of a Mott insulator based on the improved
Gutzwiller function, and show that
there is a transition from a metal to 
a charge-transfer insulator for such parameters by using the variational
Monte Carlo method.  
This transition occurs  when the level difference 
$\Delta_{dp}\equiv \epsilon_p-\epsilon_d$  between
d and p orbitals reaches a critical value $(\Delta_{dp})_c$.
The energy gain $\Delta E$, measured from the limit of large $\Delta_{dp}$,
is proportional to $1/\Delta_{dp}$ for $\Delta_{dp}>(\Delta_{dp})_c$: 
$\Delta E\propto  -t_{dp}^2/\Delta_{dp}$.
We obtain
$(\Delta_{dp})_c\simeq 2t_{dp}$ 
using the realistic band parameters. 
\end{abstract}


\maketitle


{\em Introduction}
The study of high-temperature superconductors has been 
intensively addressed since the
discovery of cuprate high-temperature superconductors.
The research of mechanism of
superconductivity (SC) in high-temperature superconductors
has attracted much attention and has been extensively studied using various 
models. 
It has been established that the Cooper pairs of high-temperature cuprates
have the $d$-wave symmetry in the hole-doped materials\cite{ben03}.
Therefore the electron correlation plays an important and the CuO$_2$
plane in cuprates plays a key role for the appearance of 
superconductivity\cite{zaa85,hir89,sca91,gue98}.
The three-band d-p model is the most fundamental model for high-temperature
cuprates\cite{hir89,sca91,gue98,koi00,yan01,yan08,yan09,web09,lau11,yan13}.

The purpose of this paper is to investigate the effect of electron
correlation in the half-filled case, that is, to discuss the
metal-insulator transition due to the on-site Coulomb repulsion.
As was discussed in Ref.\cite{zaa85}, insulators are
classified in terms of charge-transfer insulator or Mott insulator.
The cuprates belong to the class of charge-transfer insulators.
When the level difference $\Delta_{dp}\equiv\epsilon_p-\epsilon_d$ between 
$d$ and $p$ orbitals
is large, the ground state will be insulating when the Coulomb repulsion $U_d$ on
copper sites is large.  (In this paper, we use the hole picture.)
When $U_d$ is large, there will be a transition form a metal to an insulator
as $\Delta_{dp}$ is increased.  This is the Mott transition of charge-transfer type.

We will investigate this transition by using a variational Monte Carlo (VMC) method.
In correlated electron systems, we must take into account the electron
correlation correctly.
Using the VMC method we can treat the electron systems properly from weakly to 
strongly
correlated regions.
We propose a wave function for an insulator on the basis of the Gutzwiller ansatz
and examine the ground state within the space of
variational functions.
The expectation values are evaluated by using the variational Monte Carlo
algorithm\cite{gro87,yok87,nak97,yam98,yam00,miy04}.

We first discuss the Mott state of the single-band Hubbard model by proposing
a Mott-state wave function.  We show that there is a metal-insulator transition
as the on-site Coulomb repulsion $U$ is increased.  
The energy gain, compared to the limit of large $U$, is proportional to the
exchange interaction $J\equiv 4t^2/U$ in the insulating state.
The wave function is generalized  straightforwardly
to the d-p model.
In the localized region, where $\Delta_{dp}$ is greater than the critical value
$(\Delta_{dp})_c$, the energy gain $\Delta E$ is proportional to $-1/\Delta_{dp}$,
that is, $\Delta E\propto -t_{dp}^2/\Delta_{dp}$.
In this region we have the insulating ground state.
$(\Delta_{dp})_c$ is of the order of the transfer integral $t_{dp}$ between
holes in adjacent copper and oxygen atoms.  This value is consistent with
the result obtained by the dynamical mean-field theory\cite{web08}.



{\em Hamiltonian}
The three-band model that explicitly includes oxygen p and copper d orbitals 
contains
the parameters $U_d$, $U_p$, $t_{dp}$, $t_{pp}$, $\epsilon_d$ and
$\epsilon_p$.
The Hamiltonian is written as
\begin{eqnarray}
H_{dp}&=& \epsilon_d\sum_{i\sigma}d_{i\sigma}^{\dag}d_{i\sigma}
+ \epsilon_p\sum_{i\sigma}(p_{i+\hat{x}/2\sigma}^{\dag}p_{i+\hat{x}/2\sigma}
\nonumber\\
&+& p_{i+\hat{y}/2\sigma}^{\dag}p_{i+\hat{y}/2\sigma})
\nonumber\\
&+& t_{dp}\sum_{i\sigma}[d_{i\sigma}^{\dag}(p_{i+\hat{x}/2\sigma}
+p_{i+\hat{y}/2\sigma}-p_{i-\hat{x}/2\sigma}-p_{i-\hat{y}/2\sigma})\nonumber\\
&+& {\rm h.c.}]\nonumber\\
&+& t_{pp}\sum_{i\sigma}[p_{i+\hat{y}/2\sigma}^{\dag}p_{i+\hat{x}/2\sigma}
-p_{i+\hat{y}/2\sigma}^{\dag}p_{i-\hat{x}/2\sigma}\nonumber\\
&-&p_{i-\hat{y}/2\sigma}^{\dag}p_{i+\hat{x}/2\sigma}
+p_{i-\hat{y}/2\sigma}^{\dag}p_{i-\hat{x}/2\sigma}+{\rm h.c.}]\nonumber\\
&+& t_d'\sum_{\langle\langle ij\rangle\rangle\sigma}(d_{i\sigma}^{\dag}d_{j\sigma}
+{\rm h.c.} )
+ U_d\sum_i d_{i\uparrow}^{\dag}d_{i\uparrow}d_{i\downarrow}^{\dag}
d_{i\downarrow}.
\end{eqnarray}
$d_{i\sigma}$ and $d^{\dag}_{i\sigma}$ are the operators for the $d$ holes.
$p_{i\pm\hat{x}/2\sigma}$ and $p^{\dag}_{i\pm\hat{x}/2\sigma}$ denote the
operators for the $p$ holes at the site $R_{i\pm\hat{x}/2}$, and in a
similar way $p_{i\pm\hat{y}/2\sigma}$ and $p^{\dag}_{i\pm\hat{y}/2\sigma}$
are defined.  
$t_{dp}$ is the transfer integral between adjacent Cu and O orbitals
and $t_{pp}$ is that between nearest p orbitals.
$\langle\langle ij\rangle\rangle$ denotes a next nearest-neighbor pair of copper 
sites.
$U_d$ is the strength of the on-site Coulomb energy between
$d$ holes.
In this paper we neglect $U_p$ among $p$ holes because $U_p$ is small 
compared to 
$U_d$\cite{hyb89,esk89,mcm90}.
In the low-doping region, $U_p$ will be of minor importance because
the p-hole concentration is small\cite{esk91}.
The parameter values were estimated as, for example, $U_d=10.5$, $U_p=4.0$
and $U_{dp}=1.2$ in eV\cite{hyb89} where $U_{dp}$ is the nearest-neighbor
 Coulomb interaction
between holes on adjacent Cu and O orbitals.
In this paper we neglect $U_{dp}$ because $U_{dp}$ is small compared to $U_d$.
We use the notation $\Delta_{dp}=\epsilon_p-\epsilon_d$.
The number of sites is denoted as $N$, and the total number of atoms is
$N_a=3N$.
Our study is done within the hole picture where the lowest band is occupied
up to the Fermi energy $\mu$.


The single-band Hubbard model is also important in the study of strongly
correlated electron systems\cite{hub63}.  
This model is regarded as an approximation
to the three-band model. 
The Hamiltonian is given by
\begin{eqnarray}
H&=& -t\sum_{\langle ij\rangle}(c_{i\sigma}^{\dag}c_{j\sigma}+{\rm h.c.})
-t'\sum_{\langle\langle j\ell\rangle\rangle}
(c_{j\sigma}^{\dag}c_{\ell\sigma}+h.c.)\nonumber\\
&+& U\sum_in_{i\uparrow}n_{i\downarrow},
\end{eqnarray}
where $\langle ij\rangle$ and $\langle\langle j\ell\rangle\rangle$ indicate
the nearest neighbor and next-nearest neighbor pairs of sites, respectively.
$c_{i\sigma}$ and $c_{i\sigma}^{\dag}$ indicate the operators of d electrons.
$U$ is the on-site Coulomb repulsion.


{\em Band parameters and the Fermi surface}
We need an additional band parameter because we cannot reproduce the deformed
Fermi surface for cuprates by means of only $t_{dp}$ and $t_{pp}$\cite{yan14}.
Thus we have introduced the parameter $t_d'$ in the Hamiltonian in the
previous section. 
We show that the inclusion of $t_d'$ will enable us to reproduce the
curvature of the Fermi surface.
The non-zero $t_d'$ may be attributed to the integral between $d_{x^2-y^2}$ and
$d_{3z^2-r^2}$ or $d_s$ orbitals.
As we will show, the large $t_d'$ is not required to discribe the curvature
of the Fermi surface.
We must mention that there is another method to explain the curvature of
the Fermi surface.  For example, the inclusion of the O$p_x$-O$p_x$
transfer integrals leads to the deformed Fermi surface\cite{and95}.

Typical Fermi surfaces of cuprate superconductors have been reported for,
for example, (La,Sr)$_2$CuO$_4$ (LSCO) and Bi$_2$Sr$_2$CaCu$_2$O$_{8+\delta}$.
The Fermi surface for Bi$_2$Sr$_2$CaCu$_2$O$_{8+\delta}$ (Bi2212)\cite{mce03} and
Tl$_2$Ba$_2$CuO$_{6+\delta}$\cite{hus03} is deformed considerably,
while that for LSCO is likely the straight line.

For LSCO, the band parameter is estimated as $t'\sim -0.12$ when
we fit by using the single-band model.
On the other hand, Tl2201 ($T_c=93$K) and Hg1201 ($T_c=98$K) band calculations
by Singh and Pickett\cite{sin92} give very much deformed Fermi surfaces
that can be fitted by large $|t'|$ such as $t'\sim -0.4$.
For Tl2201, an Angular Magnetoresistance Oscillations (AMRO) work\cite{hus03}
gives information of the Fermi surface, which allows to get $t'\sim -0.2$
and $t''\sim 0.165$. 
There is also an Angle-Resolved Photoemission Study (ARPES)\cite{pla05}, 
which provides similar values.
In the case of Hg1201, there is an ARPES
work\cite{lee06}, from which
we obtain by fitting $t'\sim -0.2$ and $t''\sim 0.175$.

We show the Fermi surface for the d-p model in Fig.1, where
we set $t_{pp}=0$ and $t_d'=0$.
The Fermi surface shown in Fig.1 is consistent with the Fermi surface for
(La,Sr)$_2$CuO$_4$. 
However,
the deformed Fermi surfaces cannot be well fitted by using only $t_{dp}$ and
$t_{pp}$.  
We show the Fermi surface with $t_d'$ in Fig.2; the figure
indicates that the inclusion of $t_d'$ gives a deformed Fermi surface.
This Fermi surface agrees with that for Bi2212, Tl2201 and Hg1201.



\begin{figure}[htbp]
\begin{center}
\includegraphics[width=6.5cm]{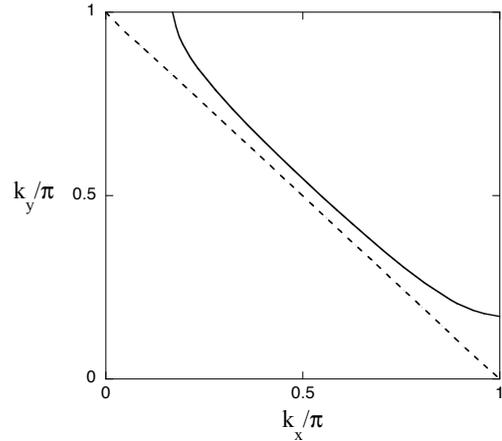}
\caption{
Fermi surface of the 2D d-p model for $t_{pp}=0$, $t_d'=0$ and $\Delta_{dp}=2.0$
in units of $t_{dp}$.  The doped-hole density is $n_h\sim 0.13$.
The dashed line is for $n_h\sim 0.0$ (half-filled case).
This Fermi surface is for LSCO.
}
\end{center}
\end{figure}



\begin{figure}[htbp]
\begin{center}
\includegraphics[width=6.8cm]{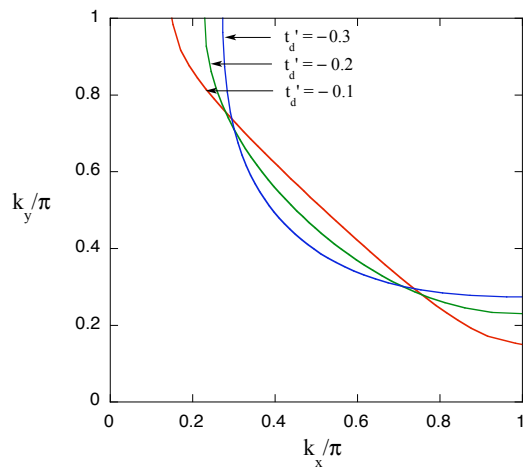}
\caption{
Fermi surface of the 2D d-p model for $t_{pp}=0.4$ and 
$\Delta_{dp}=2.0$ where $t_d'=-0.1$, $t_d'=-0.2$ and $t_d'=-0.3$
in units of $t_{dp}$.  The carrier density is $n_h\sim 0.1$.
}
\end{center}
\end{figure}


{\em Wave function of Mott state}
Here we propose a wave function to represent a Mott insulator.
We first discuss it for the single-band Hubbard model\cite{hub63}.
The energy is measured in units of $t$ in this section.
The charge-transfer Mott state of the three-band model will be given by a 
generalization of
the one-band Mott state.
The Gutzwiller function $\psi_G$ itself does not describe the insulating state
because this function has no kinetic energy gain in the limit $g\rightarrow 0$.
Wave functions for the Mott transition have been proposed for the
single-band Hubbard model by considering the doublon-holon
correlation\cite{yok04,yok06} or backflow correlations\cite{toc11}.
In the latter, a variational wave function with a Jastrow factor
is considered. 
It seems, however, not straightforward to generalize these wave functions to the
three-band case.

The Gutzwiller wave function is given as
\begin{equation}
\psi_G = P_G\psi_0 ,
\end{equation}
where $P_G$ is the Gutzwiller projection operator given by
$P_G=\prod_i[1-(1-g)n_{i\uparrow}n_{i\downarrow}]$ with the variational
parameter $g$ in the range from 0 to unity. 
$P_G$ controls the on-site electron correlation and $\psi_0$ is the Fermi sea in
this paper.
We consider the Gutzwiller function with an optimization
operator\cite{yan98}:
\begin{equation}
\psi_{\lambda} = \exp({\lambda K})\psi_G,
\end{equation}
where $K$ is the kinetic part of the Hamiltonian and $\lambda$ is a
variational parameter.
This type of wave function is an approximation to the wave function in
quantum Monte Carlo method\cite{hir81,yan07,yan13}.
The operator $e^{\lambda K}$ lowers the energy considerably.
We have finite energy gain with this function even in the limit 
$g\rightarrow 0$ due to the kinetic operator $K$.
We show that $\psi_{\lambda}$ with vanishing $g$ describes a Mott insulator.

In Fig.3 we show the energy per site as a function of $U$ on a $10\times 10$
lattice with the periodic boundary conditions.
The upper curve shows the energy for the Gutzwiller function and
the lower one is for the optimized function $\psi_{\lambda}$.
It is seen from Fig.3 that the ground-state energy changes the curvature near 
$U\sim U_c$.
This suggests that there is a transition from a metal to an insulator.
We show the parameter $g$ on $10\times 10$ lattice in Fig.4.
The parameter $g$ vanishes at a critical value $U_c\sim 8$.  
The energy for $\psi_{\lambda}$ is well approximated by a function $C/U$ with 
a constant $C$
when $U$ is large:
\begin{equation}
\frac{E}{N}\sim -C\frac{t^2}{U}\propto -CJ.
\end{equation}
This means that the energy gain mainly comes from the
exchange interaction which is of the order of $1/U$, showing that the
ground state is insulating.
The effective interaction in the limit of large $U$ is given by the
effective interaction, given by $J\sum_{\langle ij\rangle}{\bf S}_i\cdot{\bf S}_j$
with $J=4t^2/U$, and the three-site interactions\cite{har67} if we consider
only the nearest-neighbor transfer $t$.
In our calculation, we have $C\sim 3$.
This means that the ground-state energy per site is approximately given
by $E/N\sim -0.75J$.
This value, $-0.75J$, will become better by improving the wave function 
$\psi_{\lambda}$\cite{yan98}.
The inset in Fig.4 exhibits that there is a singularity in $g$ at $U\simeq U_c$ as
a function of $U$. There is a small jump in $g$.  
This indicates that the transition is first order.

The Fig.5 shows the momentum distribution function $n_k$:
\begin{equation}
n_{{\bf k}}= \frac{1}{2}\sum_{\sigma}\langle c_{{\bf k}\sigma}^{\dag}
c_{{\bf k}\sigma}\rangle.
\end{equation}
There is clearly the gap in $n_k$ at the Fermi surface for small $U$,
namely, $U\leq 6$.  In contrast, for large $U$ being greater than 7,
the gap at the Fermi surface disappears.
This indicates that the ground state is an insulator for large $U$.
This is consistent with the VMC study in Ref.\cite{yok06} where the
different trial wave function is adopted.
Other quantities are also consistent.
The critical value of $U$ is consistent; both have given $U_c\sim 7t$.
The ground-state energy is also well approximated by a curve $t^2/U$ when
$U$ is large beyond the critical value.



\begin{figure}[htbp]
\begin{center}
\includegraphics[width=6.6cm]{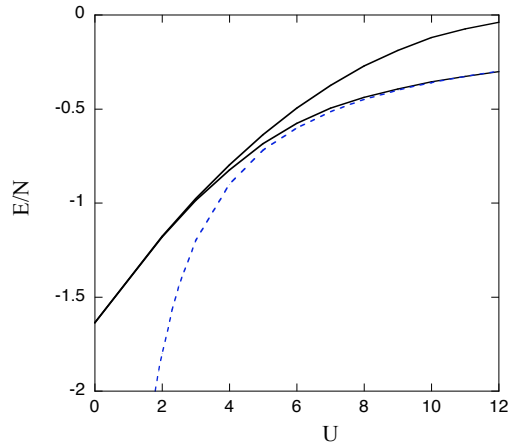}
\caption{
Ground state energy of the 2D Hubbard model per site as a function $U$ at 
half-filling on $10\times 10$ lattice.
We set $t'=-0.2$.
The upper curve is for the Gutzwiller function and the lower curve is for
$\psi_{\lambda}$.
The dotted line shows a curve C/U where $C(<0)$ is a constant.
}
\end{center}
\end{figure}

\begin{figure}[htbp]
\begin{center}
\includegraphics[angle=90,width=6.2cm]{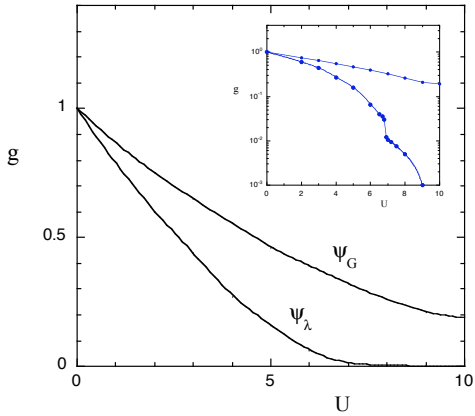}
\caption{
Gutzwiller parameter $g$ as a function of $U$ at half-filling
on $10\times 10$ lattice with $t'=-0.2$.
The parameter $g$ almost vanishes at $U\sim 8$ as for $6\times 6$ lattice.
The upper line is for the Gutzwiller function and the lower is for
$\psi_{\lambda}$.  The inset shows $g$ using a logarithmic scale.
}
\end{center}
\end{figure}

\begin{figure}[htbp]
\begin{center}
\includegraphics[width=6.8cm]{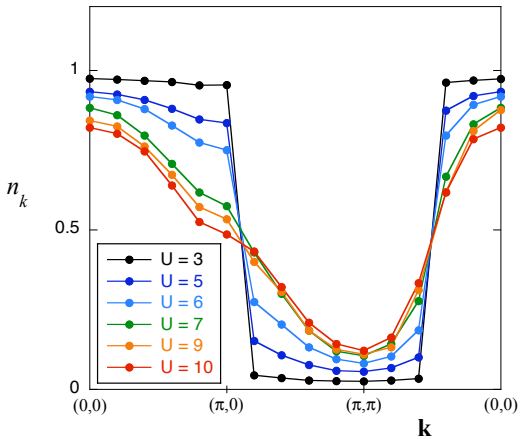}
\caption{
Momentum distribution function $n_k$ for $\psi_{\lambda}$ at half-filling
on $10\times 10$ lattice.
We show $n_k$ for $U=3,5,6,7,9$ and 10.
$n_k$ shows the insulating behavior for $U\geq 7$.
}
\end{center}
\end{figure}


\begin{figure}[htbp]
\begin{center}
\includegraphics[width=6.5cm]{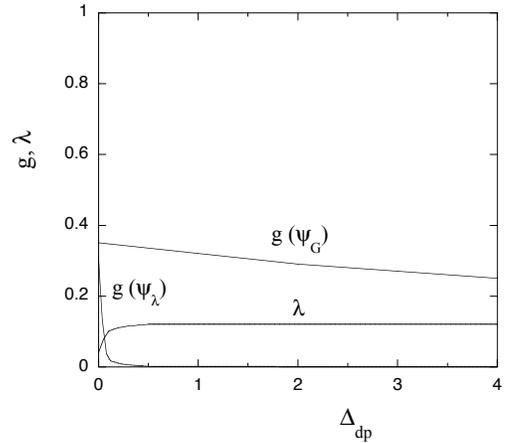}
\caption{
Parameters $g$ and $\lambda$ as a function of $\Delta_{dp}$ at half-filling
on $6\times 6$ lattice.
We used $t_{pp}=0.4$, $t_d'=-0.0$ and $U_d=8$
(in units of $t_{dp}$).
The upper line is for the Gutzwiller function and the lower is for
$\psi_{\lambda}$.
}
\end{center}
\end{figure}

\begin{figure}[htbp]
\begin{center}
\includegraphics[width=6.5cm]{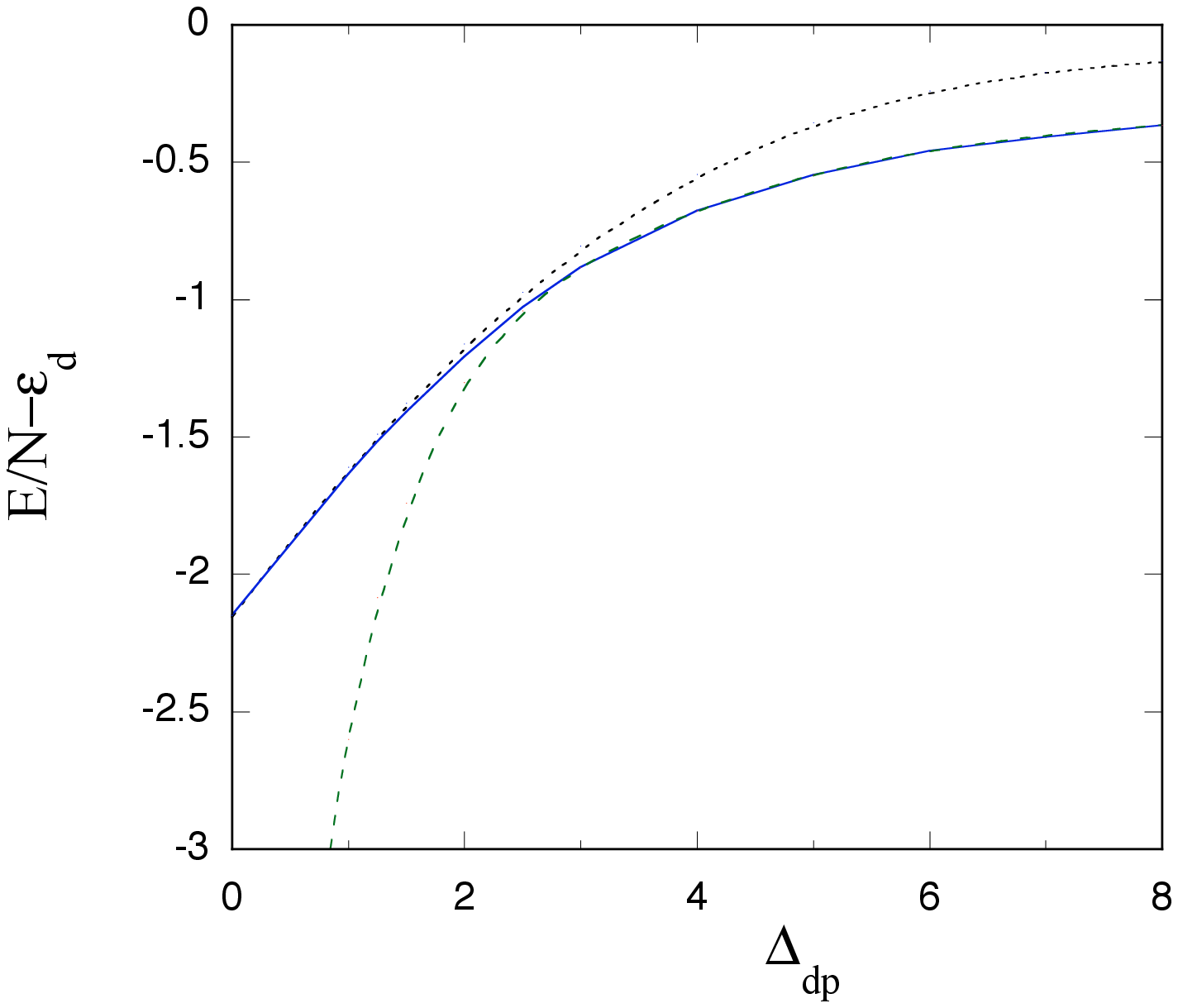}
\caption{
Ground state energy of the 2D d-p model as a function of $\Delta_{dp}$ 
for $t_{pp}=0.0$, $t_d'=-0.0$ and $U_d=8$
(in units of $t_{dp}$) in the half-filled case.
The calculations were performed on $6\times 6$ lattice.
The dotted curve is for the Gutzwiller function, namely $\lambda=0$.
The dashed curve indicates a curve given by a constant times 
$1/(\epsilon_p-\epsilon_d)$.
}
\end{center}
\end{figure}

\begin{figure}[htbp]
\begin{center}
\includegraphics[width=6.5cm]{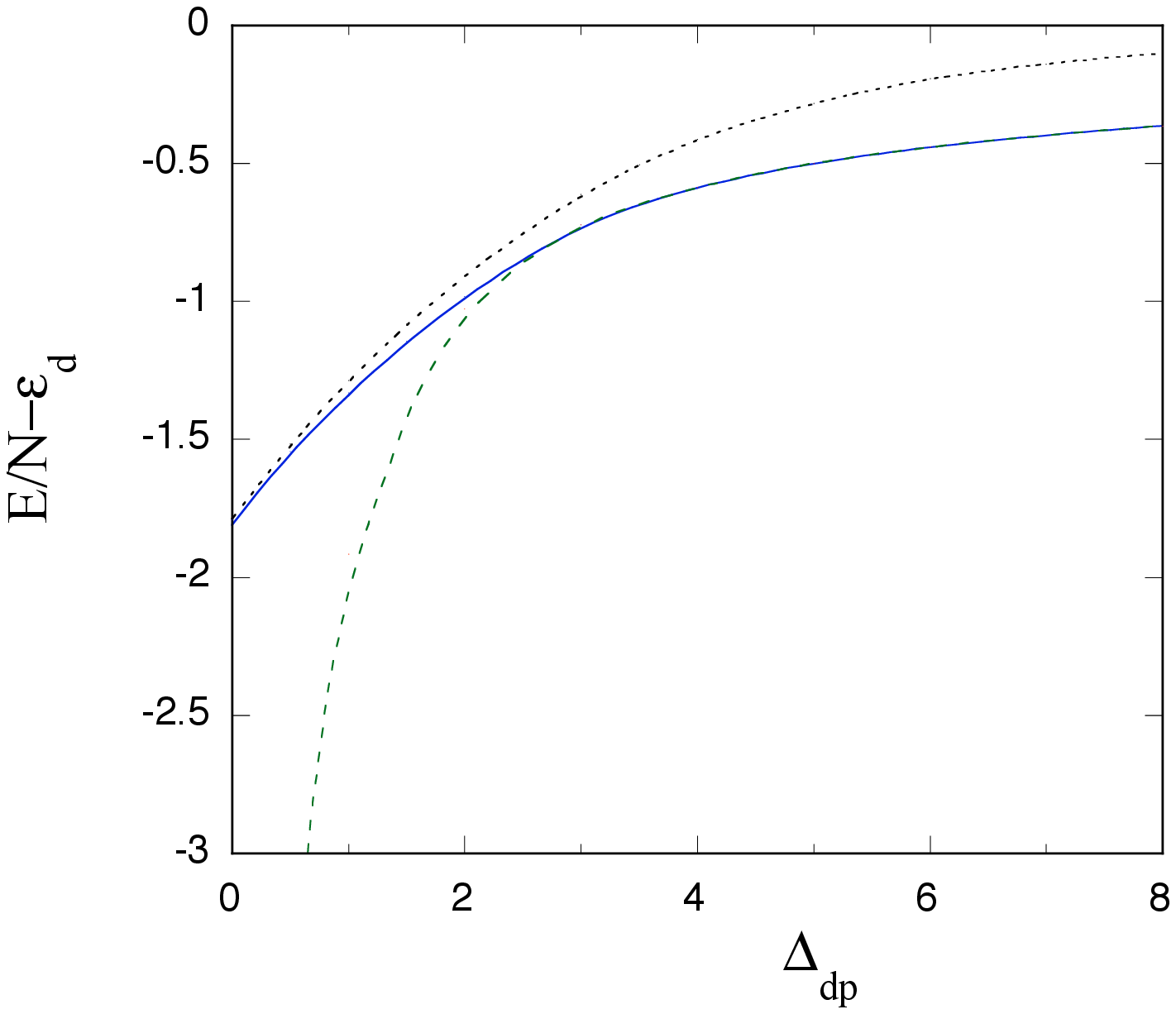}
\caption{
Ground state energy of the 2D d-p model as a function of $\Delta_{dp}$ 
for $t_{pp}=0.4$, $t_d'=-0.2$ and $U_d=8$
(in units of $t_{dp}$) in the half-filled case.
The calculations were performed on $6\times 6$ lattice.
The dotted curve is for the Gutzwiller function with $\lambda=0$.
The dashed curve indicates a curve given by a constant times 
$1/(\epsilon_p-\epsilon_d)$.
}
\end{center}
\end{figure}

\begin{figure}
\begin{center}
\includegraphics[width=6.5cm]{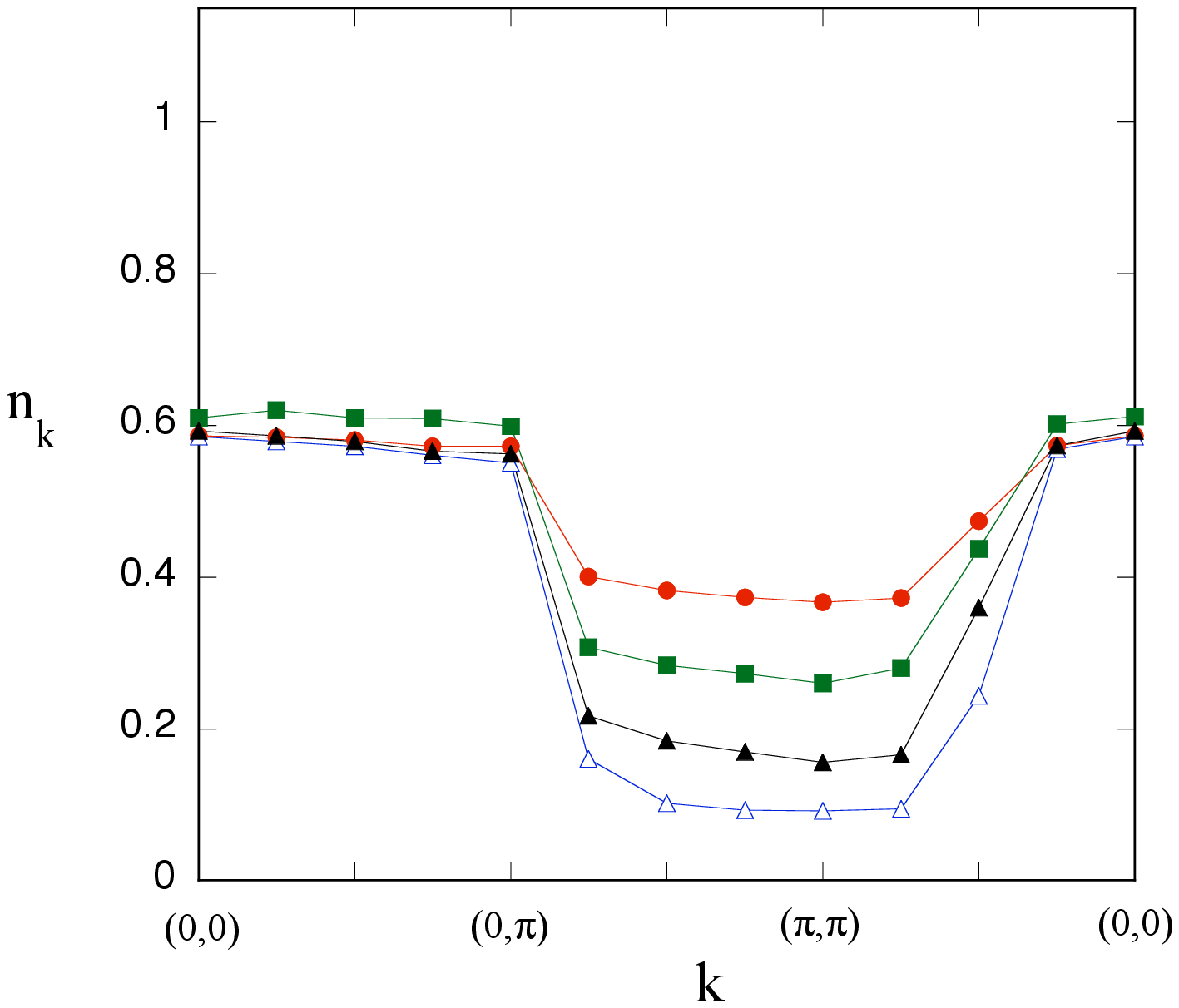}
\caption{
Momentum distribution function of d holes in the 2D d-p model
for $t_{pp}=0.4$, $t_d'=-0.0$ and $U_d=8$
(in units of $t_{dp}$) in the half-filled case
on $8\times 8$ lattice.
From the top, $\epsilon_p-\epsilon_d=6$, 4, 2 for $\psi_{\lambda}$ and
the bottom is for the Gutzwiller function with $\epsilon_p-\epsilon_d=2$.
}
\end{center}
\end{figure}



{\em Charge-transfer Mott state}
In this section, we consider the ground state of the three-band d-p model 
in the half-filled case.
The energy unit is given by $t_{dp}$ in this section.
The Gutzwiller wave function for the d-p model is
$\psi_G = P_G\psi_0$,
where $P_G$ is the Gutzwiller projection operator for d electrons.
We neglect the on-site Coulomb repulsion on the oxygen site because it is
not important when the number of p holes is small.
$\psi_0$ is a one-particle wave function given by the Fermi sea. 
$\psi_0$ contains the variational parameters $\tilde{t}_{dp}$,
$\tilde{t}_{pp}$, $\tilde{t}_d'$, $\tilde{\epsilon}_d$ and 
$\tilde{\epsilon}_p$:
\begin{equation}
\psi_0= \psi_0(\tilde{t}_{dp}, \tilde{t}_{pp}, \tilde{t}_{d}',
\tilde{\epsilon}_d, \tilde{\epsilon}_p).
\end{equation}
In the non-interacting case, $\tilde{t}_{dp}$, $\tilde{t}_{pp}$ and
$\tilde{t}_{d}'$ coincide with $t_{dp}$, $t_{pp}$ and $t_d'$ in the
Hamiltonian, respectively.
As we have shown, $t_d'$ plays an important role to determine the Fermi
surface within the d-p model.
The Fermi surface is determined by
$\tilde{t}_{dp}$, $\tilde{t}_{pp}$ and $\tilde{t}_{d}'$ in the 
correlated wave function.

The optimized wave function is
\begin{equation}
\psi_{\lambda} = \exp(\lambda K)\psi_G,
\end{equation}
where $K$ is the kinetic part of the total Hamiltonian $H_{dp}$ and
$\lambda$ is a variational parameter.
In general, the band parameters $t_{pp}$, $t'_d$, $\epsilon_d$ and 
$\epsilon_p$ in $K$ are regarded as variational parameters: 
\begin{equation}
K = K(\hat{t}_{pp},\hat{t'_d},\hat{\epsilon_d},
\hat{\epsilon_p}).
\end{equation}
For simplicity, we take $\hat{t}_{pp}=\tilde{t}_{pp}$, $\hat{t'}_d=\tilde{t'}_d$,
$\hat{\epsilon}_{d}=\tilde{\epsilon}_{d}$ and $\hat{\epsilon}_p=\tilde{\epsilon}_p$.
The energy expectation value is minimized for variational parameters $g$,
$\tilde{t}_{dp}$, $\tilde{t}_{pp}$, $\tilde{t}_{d}'$,
$\tilde{\epsilon}_p-\tilde{\epsilon}_d$ and $\lambda$.


We show the parameter $g$ as a function of $\Delta_{dp}$ for $U_d=8$, $t_{pp}=0.4$
and $t'_d=0$ on $6\times 6$ lattice in Fig.6.
$g$ of $\psi_{\lambda}$ is decreasing rapidly for positive $\Delta_{dp}$ and 
vanishes at $\Delta_{dp}\sim 0.5t_{dp}$ while that
in $\psi_G$ decreases gradually as a function of $\Delta_{dp}$.
The figure 8 exhibits the ground-state energy per site 
$\Delta E\equiv E/N-\epsilon_d$ 
as a function of $\Delta_{dp}$ for $t_{pp}=0$, $t_d'=0$ and $U_d=8$.
This set of parameters correspond to that for LSCO.
We can find that the curvature of the energy, as a function of $\Delta_{dp}$,
is changed near $\Delta_{dp}\sim 2t_{dp}$.
This means that the region $\Delta_{dp} > 2t_{dp}$ is a large-$\Delta_{dp}$ region.
The energy is well fitted by $1/\Delta_{dp}$ shown by the dashed curve
in Fig.7 when $\Delta_{dp}$ is
greater than the critical value $(\Delta_{dp})_c\sim 2t_{dp}$.
A similar behavior is also observed for the other set of parameters such as
$t_{pp}=0.4$, $t_d'=-0.2$ and $U_d=8$ for Bi2212, Tl2201 and Hg1201,
as shown in Fig.8.
We show the momentum distribution for d electrons defined by
$n_k=\langle d_{k\sigma}^{\dag}d_{k\sigma}\rangle$ in Fig.9.
This was calculated on $8\times 8$ lattice and exhibits the effect of 
correlation in the localized region.

The results indicate that the energy gain $\Delta E$ mostly comes from the 
second-order perturbation with the excitation energy $\Delta_{dp}$.
Thus, in general, the energy gain $\Delta E$ is expanded in terms of
$\Delta_{dp}^{-1}$ when $\Delta_{dp}$ is large:
\begin{equation}
\Delta E \simeq \frac{A_1}{\Delta_{dp}}+\frac{A_2}{\Delta_{dp}^2}+\cdots.
\end{equation}
As seen from Figs.7 and 8, $A_1$ is negative and $A_2$ is positive.
It is known that the antiferromagnetic exchange interaction works between
d electrons on neighboring copper atoms.  The coupling $J_{Cu-Cu}$ is
given as\cite{esk93,kam94}
\begin{equation}
J_{Cu-Cu}= \frac{4t_{dp}^2}{(\Delta_{dp}+U_{dp})^2}\left( \frac{1}{U_d}
+\frac{2}{2\Delta_{dp}+U_p} \right).
\end{equation}
The exchange coupling $J_{Cu-Cu}$ will give the energy gain being
proportional to $1/\Delta_{dp}^2$ when the d electrons are antiferromagnetically
aligned on copper atoms.  Our results show that this contribution is
small keeping $A_2$ positive.

For large $\Delta_{dp}$ the energy gain is proportional to $1/\Delta_{dp}$:
\begin{equation}
\Delta E=\frac{E}{N}-\epsilon_d \simeq -C\frac{t_{dp}^2}{\Delta_{dp}},
\end{equation}
for a constant $C$.
This indicates
that the ground state is an insulator of charge-transfer type.
In our calculation we obtain $C\sim 3$.


{\em Summary}
We have proposed a wave function of Mott insulator based on an optimized
Gutzwiller function in strongly correlated electron systems.
We have investigated Mott transition at half-filling in the single-band
Hubbard model first and generalized it to the three-band
d-p model by employing the variational Monte Carlo method.
The metal-insulator transition occurs as a result of strong correlation.
The wave function in this paper describes a first-order transition from a metal
to a Mott insulator.

Our wave function has the form
\begin{equation}
\psi_{\lambda} = e^{\lambda K}\psi_G(g),
\end{equation}
where $g$ is the Gutzwiller parameter.
The limit $g\rightarrow 0$ indicates no double occupancy of d holes 
in $\psi_G$.
In this limit the energy of the single-band Hubbard model is given by
that of the strong-coupling limit $U\gg t$, namely, $E\propto -t^2/U$.
This means that $\psi_{\lambda}$ is an insulator in the limit 
$g\rightarrow 0$.  This state indeed becomes stable when $U$ is as large
as $7t$.  This shows that there is a metal-insulator transition at
$U=U_c\sim 7t$.
There is a singularity in $g$ at $U\simeq U_c$ as a function of $U$,
indicating that the transition is first order.
The same discussion also holds for the three-band d-p model except that
the cuprates exhibit charge-transfer transition.
In the localized region with large $\Delta_{dp}$, the energy of $\psi_{\lambda}$
with vanishing $g$ is given by $E/N-\epsilon_d\sim -t_{dp}^2/\Delta_{dp}$, 
indicating that
$\psi_{\lambda}$ is a charge-transfer insulator.
The stabilization of $\psi_{\lambda}$ for large $U_d$ and $\Delta_{dp}$
shows the existence of a metal-insulator transition in the d-p model. 
The transition occurs for the band parameters that are suitable
for high temperature cuprates.
Our result shows that $(\Delta_{dp})_c\sim 2t_{dp}$.
If we use $t_{dp}\simeq 1.5$eV\cite{hyb89,esk89,mcm90}, the charge-transfer
insulator has a gap of 3eV.
 
Finally, we give a discussion on magnetism.
The competition between magnetic state and paramagnetic state would depend 
on band parameters.  
There would be a transition from a magnetic insulator to a paramagnetic
insulator as the band parameters are varied.
We expect that $t_d'$ plays an important role in this transition
because $t_d'$ would play a similar role to the next-nearest neighbor
transfer integral $t'$ in the single-band Hubbard model.

We express sincere thanks to J. Kondo, K. Yamaji and I. Hase for useful 
discussions.
This work was supported by a Grant-in-Aid for Scientific Research
from the Ministry of Education, Culture, Sports, Science and Technology
of Japan.
A part of the numerical calculations was performed at the Supercomputer
Center of the Institute for Solid State Physics, University of Tokyo.

\end{document}